\documentclass[aps,prd,amsmath,amssymb,nofootinbib,showpacs,reprint]{revtex4-1}

\usepackage{etex}
\usepackage{amssymb,amsmath,amsfonts,amsthm}
\usepackage{graphicx}
\usepackage{slashed}
\usepackage{bbm,bm}
\usepackage{mathtools}
\usepackage{color,array,tabularx}
\usepackage{stackrel}
\usepackage[small,bf]{caption}
\usepackage{pst-all}
\usepackage{psfrag}
\usepackage{pstricks,pst-grad,color,shadow}
\usepackage{tikz}
\usetikzlibrary{shapes,arrows,backgrounds}
\usepackage{soul}
\usepackage{placeins}
\usepackage{latexsym}
\usepackage{keyval}
\usepackage{ifthen}
\usepackage{moreverb}
\usepackage{gnuplottex}
\usepackage{simplewick}
\usepackage{epigraph}
\usepackage{color}
\usepackage{graphicx}
\usepackage{bm}
\usepackage{bbm}
\usepackage{pst-grad}

\usepackage{amsfonts}
\usepackage{amsmath}
\usepackage{amssymb}
\usepackage{amsthm}

\usepackage{psfrag}
\usepackage{pstricks,pst-grad,color,shadow}
\usepackage{dsfont}
\usepackage{tikz}
\usetikzlibrary{shapes,arrows,backgrounds}
\usepackage{soul}
\usepackage{placeins}
\usepackage{wrapfig}

\definecolor{darkgreen}{rgb}{0,0.73,0}

\newcommand{\erw}[1]{\left \langle #1 \right \rangle}

\newcommand{\algebra}[1]{\mathfrak{#1}}
\newcommand{\group}[1]{\mathcal{#1}}

\newcommand{\gU}{\mathcal{U}}

\newcommand{\trnsp}{\mathsf{T}}

\DeclareMathOperator{\tr}{tr}

\renewcommand{\Re}{\operatorname{Re}}

\newcommand{\ii}{\mathrm{i}}

\newcommand{\norm}[1]{\left \vert \left \vert #1 \right \vert \right \vert}

\newcommand{\Rep}[1]{(#1)}
\newcommand{\chargec}{\mathsf{C}}

\renewcommand{\Re}{\operatorname{Re}}

\newcommand{\egamma}{{\gamma_\mathsf{E}}}

\graphicspath{{./}{./plots/}{./publPlots/}}

\begin{document}
\title{Hadron masses and baryonic scales in $G_2$-QCD at finite density}

\newcommand{\TUD}{Theoriezentrum, Institut f\"ur Kernphysik, Technische Universit\"at Darmstadt, 64289 Darmstadt, Germany} 
\newcommand{\FSU}{Theoretisch-Physikalisches Institut, Friedrich-Schiller-Universit{\"a}t Jena, 
07743 Jena, Germany}
\newcommand{\JLU}{Institut f\"ur Theoretische Physik, Justus-Liebig-Universit\"at Giessen, 35392 Giessen, Germany}

\author{Bj\"orn H. Wellegehausen}
\email{Bjoern.Wellegehausen@uni-jena.de}
\affiliation{\JLU}
\affiliation{\FSU}

\author{Axel Maas}
\email{axelmaas@web.de}
\affiliation{\FSU}

\author{Andreas Wipf}
\email{Wipf@tpi.uni-jena.de}
\affiliation{\FSU}

\author{Lorenz von Smekal}
\email{lorenz.smekal@physik.tu-darmstadt.de}
\affiliation{\JLU} 
\affiliation{\TUD}

\begin{abstract}
\noindent The QCD phase diagram at densities relevant to neutron stars remains elusive, mainly due to the fermion-sign problem. 
At the same time, a plethora of possible phases has been predicted in models. 
Meanwhile $G_2$-QCD, for which the $SU(3)$ gauge group of QCD is replaced by the exceptional Lie group $G_2$, does not have a sign problem and can be simulated at such densities using standard lattice techniques. 
It thus provides benchmarks to models and functional continuum methods, and it serves to unravel the nature of possible phases of strongly interacting matter at high densities. 
Instrumental in understanding these phases is that $G_2$-QCD has fermionic baryons, and that it can therefore sustain a baryonic Fermi surface. 
Because the baryon spectrum of $G_2$-QCD also contains bosonic diquark and probably other more exotic states, it is important to understand this spectrum before one can disentangle the corresponding contributions to the baryon density. 
Here we present the first systematic study of this spectrum from lattice simulations at different quark masses. 
This allows us to relate the mass hierarchy, ranging from scalar would-be-Goldstone bosons and intermediate vector bosons to the $G_2$-nucleons and deltas, to individual structures observed in the total baryon density at finite chemical potential. 
\end{abstract}


\maketitle

\section{Introduction}

\noindent Understanding neutron stars requires to understand the equilibrium properties of nuclear and hadronic matter \cite{Leupold:2011zz,BraunMunzinger:2009zz} at low temperature and high density. 
This is essential in every stage starting from neutron star formation and cooling to neutron star mergers, and hence ultimately to understanding the synthesis of the heavy elements. 
A serious technical obstacle in this process is the infamous fermion-sign problem, which prevents efficient numerical simulations of the underlying theory of nuclei and hadrons, QCD \cite{Gattringer:2010zz,deForcrand:2010ys,Philipsen:2011zx}. 
Although substantial progress has been achieved with models and functional  
continuum methods \cite{Leupold:2011zz,Buballa:2003qv,Pawlowski:2010ht,Braun:2011pp}, input from lattice simulations remains indispensable.   

There have been several approaches to circumvent the sign problem, e.g.\ analytic continuations from imaginary \cite{Bonati:2012pe,deForcrand:2010he,Cea:2012ev} or isospin \cite{Kogut:2004zg,deForcrand:2007uz} chemical potential which fail, however, when phase transitions are encountered. 
Another possibility is to combine strong-coupling and hopping expansion techniques to derive an effective theory for heavy quarks \cite{Fromm:2011qi,Fromm:2012eb} whose range of applicability must then be assessed. 
Further alternatives might be provided by stochastic approaches \cite{Sexty:2013ica}, but it is as yet unclear whether they will eventually solve the problem in QCD.  

A complementary strategy is to use QCD-like theories without a sign problem. This strategy serves two aims. 
One is to provide numerical benchmarks for model building \cite{Leupold:2011zz,Buballa:2003qv} and continuum methods \cite{Pawlowski:2010ht,Braun:2011pp}, for continuations from imaginary or isospin chemical potential, and equally so for the effective lattice theories for heavy quarks.  
The other is to gain insight into the genuine properties of gauge theories  
other than QCD at finite densities, and to exploit analogies with other physical systems such as ultracold fermionic quantum gases. 
Such QCD-like theories include two-color QCD \cite{Kogut:2000ek,Hands:2000ei,Hands:2006ve,Hands:2011ye,Strodthoff:2011tz,Strodthoff:2013cua,Boz:2013rca} and adjoint QCD \cite{Kogut:2000ek,Karsch:1998qj,Engels:2005te,Bilgici:2009jy}. 
However, neither of these directly compare well with QCD. Two-color QCD with fundamental quarks does not have fermionic baryons \cite{Kogut:2000ek,Hands:2000ei}, while adjoint QCD is known to behave rather differently from QCD already in the quenched case \cite{Danzer:2008bk}.

We have recently added another such replacement theory, $G_2$-QCD \cite{Holland:2003jy,Maas:2012wr}, and shown that it is possible to simulate this theory at finite density and temperature. 
This permitted a first view of the full phase diagram of $G_2$-QCD. We will discuss the properties of this theory in detail in Section \ref{sg2props}. 
Here, it suffices to state that it can be simulated without fermion-sign problem at finite density, it does have fermionic baryons, and its properties in the quenched case are very similar to QCD as well. 
Especially this last observation has quite interesting implications for the role of the center symmetry in QCD. 
A brief review and guide to the literature 
is given in \cite{Maas:2012ts}. 

In order to better understand the physical picture behind the phase diagram of this theory, however, one needs to understand its hadronic spectrum. 
In \cite{Maas:2012wr} we studied a few low-lying states to give a rough estimate of the scales involved in the simulations. 
To firmly identify the properties of various finite densisty phases, we need a much clearer picture of the hadron masses and the corresponding hierarchy of mass scales. 
These can be deduced from the spectrum of hadronic states in the vacuum. To determine this spectrum from lattice Monte-Carlo simulations is the main purpose of the present article. 
We discuss the theoretical foundations of (lattice) spectroscopy for $G_2$-QCD in Section \ref{sspectheo}. 
While the lattice determination of the spectrum is in principle straightforward, it is a rather challenging task, when it comes to the details which we describe  in Section \ref{stech}. 
The results for spectra obtained with two different quark masses are presented in Section~\ref{sspectrum}.  

To show that this information is indeed relevant for understanding the phase diagram we relate these results in an explorative way to the dependence of the quark density on their chemical potential in Section \ref{sdensity}. 
We thereby observe various structures corresponding to the hierarchy of scales in the spectrum given by the baryon masses per quark number. 
Especially, we find an onset at half the would-be-Goldstone mass, a stepwise increase in density at half the intermediate vector boson mass, and a rapid further growth setting in at around one third of the nucleons' mass which is characteristic of their fermionic nature and which might be a manifestation of $G_2$ nuclear matter. 
The results indeed suggest that the theory has a rich phase structure, and that baryon-dominated regions of the phase diagram exist before the density is eventually dominated by quarks and lattice artifacts at large chemical potentials. 
This is of significant importance, as it might indeed point towards the presence of a baryonic Fermi surface, making $G_2$-QCD a viable model to understand generic features of the finite density phases of the strong interaction.

Our results are summarized once more together with our conclusions in Section \ref{sconclusions}. Note that some preliminary material was already presented in \cite{Maas:2012ts}.  

\section{General properties of $G_2$-QCD}\label{sg2props}

\noindent The action of $N_\text{f}$ flavour QCD with arbitrary gauge group $\group{G}$ in Minkowski space-time is given by
\begin{equation}
\begin{aligned}
S=&\int d^4 x \tr \left \lbrace-\frac{1}{4} F_{\mu\nu}F^{\mu\nu} + \right .\\
&\left . \sum \limits_{n=1}^{N_\text{f}} \bar{\Psi}_n\left(\ii \, \gamma^\mu
(\partial_\mu-g A_\mu)-m \right)\Psi_n\right \rbrace,
\label{eqn:actionQCD}
\end{aligned}
\end{equation}
with $A_\mu$ an element of the corresponding gauge algebra $\algebra{g}$. 
For QCD, the gauge group is SU(3), but here we will use instead the
exceptional Lie group $G_2$.  
For the sake of completeness, we will briefly review the construction
of the gauge group $G_2$ in Section \ref{sg2}, reviewing parts of
Ref.~\cite{Holland:2003jy}, before we turn towards the quark sector.   
The most important ingredient is the Dirac operator, to be discussed
in Section \ref{sdirac}, and the realization of chiral symmetry
discussed in Section \ref{schiral}.  
Because $G_2$ is a real group, chiral symmetry breaking and the
concept of baryon number require special attention, as described in
Section \ref{sbn}. 

\subsection{Construction of the gauge group $G_2$}\label{sg2}

\noindent
$G_2$ is the smallest of the five exceptional simple Lie groups and it is
also the smallest simple and simply connected Lie group which has a trivial
center. As $SU(3)$, the gauge group of the strong interactions, it has rank $2$. The fundamental representations are $7$-dimensional and 
$14$-dimensional, the latter coinciding with the adjoint
representation. The elements of $G_2$ can be viewed as 
elements of $SO(7)$ subject to seven independent cubic constraints for the
$7$-dimensional matrices representing the Lie algebra of $SO(7)$
\cite{Borowiec:2004tv,Holland:2003jy},
\begin{equation}
\label{eq:g2constraint}
T_{abc} = T_{def}\,g_{da}\,g_{eb}\,g_{fc},
\end{equation}
where $T$ is a totally antisymmetric tensor. There are thus
$N_\mathsf{c}=7$ quark colors and 14 gluons in $G_2$. 

The constraints \eqref{eq:g2constraint} reduce the number of
generators from $21$ for $SO(7)$ to
$14$ for the group $G_2$. In addition, $G_2$ is connected to $SU(3)$
through the embedding of $SU(3)$ as a subgroup of $G_2$ according to
\cite{Macfarlane:2002hr,2009arXiv0902.0431Y}
\begin{equation}
G_2/SU(3) \sim SO(7)/SO(6) \sim S^6.\label{cosetG2}
\end{equation}
This means that every element $\gU$ of $G_2$ can be written as
\begin{equation}
\begin{aligned}
\gU=\group{S} \cdot \group{V} \quad \text{with} & \quad\group{S} \in
G_2/SU(3) \\ 
\text{and} & \quad \group{V} \in SU(3).\label{decomposition}
\end{aligned}
\end{equation}
In the pure $G_2$ gauge theory
\cite{Pepe:2006er,Wellegehausen:2011sc,Maas:2007af} this decomposition
is in fact being used to speed up the numerical simulations.
Since $G_2$ is a subgroup of $SO(7)$, all representations are real and one can
always choose a real basis for the Lie algebra. A possible real
representation for the 14 generators is given explicitly in
Refs.~\cite{Cacciatori:2005yb,Greensite:2006sm}. 

\subsection{The spectrum of the Dirac operator}\label{sdirac}

\noindent
For lattice Monte-Carlo methods to be applicable, the determinant of
the Euclidean Dirac 
operator has to be non-negative. The continuum Dirac operator is given by
\begin{equation}
D[A,m,\mu]=\egamma^\mu (\partial_\mu-g A_\mu)-m +\egamma_0 \mu.
\end{equation}
where the Euclidean gamma matrices are Hermitian. As in QCD it satisfies
\begin{equation}
D(\mu)^\dagger \,\gamma_5=\gamma_5\,D(-\mu^*)
\end{equation}
and the fermion determinant is real at imaginary chemical potential.
In addition, however, the $G_2$ Dirac operator also satisfies the relation 
\begin{equation}
\begin{aligned}
D(\mu)^* \,T=T\,D(\mu^*)\,\quad \text{with} \\ T=C\gamma_5,\quad T^*\, T=-\mathbbm{1}
,\quad T^\dagger=T^{-1}, \label{eqRealityCond}
\end{aligned}
\end{equation}
where $C$ is charge conjugation matrix.
If such a unitary operator $T$ exists then the
eigenvalues of the Dirac operator come in complex conjugate pairs and all real
eigenvalues are doubly degenerate 
\cite{Kogut:2000ek,Hands:2000ei}, analogous to the Kramers degeneracy
of time-reversal invariant spin Hamiltonians. Therefore
\begin{equation}
\det D[\,A,m,\mu]\geq 0 \quad \text{for} \quad \mu \in \mathbbm{R}.
\end{equation}
This property of the fermion determinant makes Markov chain Monte-Carlo
techniques applicable even at finite densities, because the path integral measure $\mathcal{D}A_\mu \det
D[\,A,m,\mu]\, e^{-S_\mathsf{B}}$ then essentially provides 
 a probability distribution.

\subsection{Chiral symmetry}\label{schiral}

\noindent
In \cite{Kogut:2000ek}, the chiral symmetry of different gauge groups
has been investigated. 
Here we review the details for $G_2$, see also \cite{Holland:2003jy}. Under charge conjugation the matter part of the Lagrange density transforms, up
to boundary terms, as
\begin{equation}
\begin{aligned}
\mathcal{L}[
\Psi^\chargec,A,m]=\mathcal{L}[\Psi,-A^T,m],
\end{aligned}
\end{equation}
with $\Psi=\left(\Psi_1, \dots,\Psi_{N_\text{f}}\right)$. Therefore, the charge
conjugated spinor $\Psi^\chargec$ fulfills the same equations of motion 
as $\Psi$ if the gauge field obeys the condition
\begin{equation}
A_\mu^\trnsp=-A_\mu=-A_\mu^a T_a.
\label{eqn:majCondA}
\end{equation}
Since every representation of $G_2$ is real, the generators $T_a$ of the algebra
$\algebra{g}_2$ can be chosen as anti-symmetric real-valued $7\times7$
matrices and hence Equation (\ref{eqn:majCondA}) holds.

It is then possible to write the matter part of the action \eqref{eqn:actionQCD} as a sum over
$2N_\mathsf{f}$ Majorana spinors $\lambda_n$
\begin{equation}
\begin{aligned}
\mathcal{L}[\Psi,A]=&\bar{\Psi}\left(\ii \, \gamma^\mu
(\partial_\mu-g A_\mu)-m \right)\Psi\\
=&\bar{\lambda}\left(\ii \,
\gamma^\mu (\partial_\mu-g A_\mu)-m\right)\lambda
\label{eqn:actionQCDMaj}
\end{aligned}
\end{equation}
with $\lambda=(\chi\,,\eta)=(\lambda_1,\dots,\lambda_{2N_\text{f}})$. Here
$\lambda$ obeys the Majorana condition $
\lambda^\chargec=C \bar{\lambda}^\trnsp=\lambda$, $
\bar{\lambda}^\chargec=-\lambda^\trnsp C^{-1}=\bar{\lambda}$,
and it is related to the Dirac spinor as
\begin{equation}
\begin{aligned}
\Psi=&\chi+\ii\, \eta\,,\quad\bar{\Psi}=\bar{\chi}-\ii\, \bar{\eta}\,,\\
\Psi^\chargec=&\chi-\ii\, \eta\,,\quad \quad\bar{\Psi}^\chargec=\bar{\chi}+\ii\,
\bar{\eta}.
\end{aligned}
\end{equation}
Therefore, it follows that $G_2$-QCD possesses an extended flavour
symmetry as compared to SU(3)-QCD. 

The action is invariant under the $SO(2N_\mathsf{f})_\mathsf{V}$
vector transformations 
\begin{equation}
\lambda \mapsto  e^{\beta \otimes \mathbbm{1}}\lambda
\end{equation}
with a real and antisymmetric $\beta\in\algebra{so}(2 N_\mathsf{f})$, 
and under the axial transformations
\begin{equation}
 \lambda \mapsto e^{\ii \, \alpha \otimes \gamma_5}\lambda
\end{equation}
with a real symmetric matrix $\alpha$. These do not form a group,
but the transformations with diagonal $\alpha$ form the group
$U(1)^{2N_\mathsf{f}}$ and those with $\alpha\propto \mathbbm{1}$
among them generate the axial $U(1)$.
Due to the Majorana constraint left- and right-handed spinors cannot be rotated independently. 
The general transformation is a composition of an
axial- and  a vector transformation,
\begin{equation}
\begin{aligned}
\lambda &\mapsto  e^{\beta\otimes\mathbbm{1}}e^{\ii \alpha\otimes\gamma_5} \lambda
\equiv V(\alpha,\beta)\lambda\,\\ 
V&=
U(\alpha,\beta) \otimes P_\mathsf{L}+
U^*(\alpha,\beta)\otimes
P_\mathsf{R},
\end{aligned}
\end{equation}
with an $U(2N_\mathsf{f})$-matrix $U(\alpha,\beta)=
e^{\beta}e^{\ii\alpha}$, in agreement with the results in \cite{Holland:2003jy}.
Following the same arguments as in QCD it is expected that 
the axial $U(1)$ is broken by the axial anomaly such 
that only an extended $SU(2N_\mathsf{f})\times \mathbbm{Z}(2)_\mathsf{B}$
chiral symmetry remains.

\subsection{Chiral symmetry breaking and baryon number}\label{sbn}
In the presence of a non-vanishing \emph{Dirac mass term} (or a non-vanishing chiral
condensate) the theory is no longer invariant under the axial
transformations. Therefore the non-anomalous chiral symmetry is
expected to be broken explicitly (or spontaneously) to its
maximal vector subgroup,
\begin{equation}
\begin{aligned}
SU({2N_\mathsf{f}})\otimes
\mathbbm{Z}(2)_\mathsf{B} \,\stackrel{m}{\mapsto}\,
SO(2 N_\mathsf{f})_\mathsf{V}\otimes \mathbbm{Z}(2)_\mathsf{B},
\end{aligned}
\end{equation}
leading to
$N_\mathsf{f}(2N_\mathsf{f}+1)-1$
(would-be) Goldstone bosons. 

\noindent
The (baryon) \emph{chemical potential} for a Dirac fermion enters the partition function as an
off-diagonal term in Majorana flavor space,
\begin{equation}
\begin{aligned}
\mathcal{L}=&\bar{\Psi} \left(\ii\, \slashed{D}-m + \ii \,\gamma_0\,
\mu\right)\Psi \\ 
=&\begin{pmatrix} \bar{\chi} \\ \bar{\eta}\end{pmatrix}
\begin{pmatrix} \ii\,\slashed{D}-m & \ii\,\gamma_0\, \mu\\ -\ii\,\gamma_0\, \mu & \ii\,\slashed{D}-m \end{pmatrix} \begin{pmatrix} \chi \\
\eta\end{pmatrix}.
\label{eqn:lagrangeQCDchem}
\end{aligned}
\end{equation}
With chemical potential but vanishing Dirac mass the 
remaining chiral symmetry is thus the same as in QCD,
\begin{equation}
\begin{aligned}
SU({2N_\mathsf{f}})&\otimes \mathbbm{Z}(2)_\mathsf{B} \stackrel{\mu}{\mapsto} \\
SU(N_\mathsf{f})_\mathsf{A}&\otimes
SU(N_\mathsf{f})_\mathsf{V} \otimes U(1)_\mathsf{B}/\mathbbm{Z}(N_\mathsf{f}).
\end{aligned}
\end{equation}
For $m\neq 0$ the remaining chiral symmetry is further broken as
\begin{equation}
\begin{aligned}
SU(N_\mathsf{f})_\mathsf{A}\otimes
SU(N_\mathsf{f})_\mathsf{V} \otimes U(1)_\mathsf{B}/\mathbbm{Z}(N_\mathsf{f})
\stackrel{\mu,m}{\mapsto} \\
SU(N_\mathsf{f})_\mathsf{V} \otimes U(1)_\mathsf{B}/\mathbbm{Z}(N_\mathsf{f}).
\end{aligned}
\end{equation}
If one first introduces a mass and only afterwards a 
chemical potential then one notices, that for $\mu\neq 0$ the Lagrangian
is off-diagonal in the Majorana basis such that is not possible to transform 
the Majorana components of a Dirac spinor independently. 
Therefore, the vector symmetry $SO(2N_\mathsf{f})_\mathsf{V}$ 
of the massive theory is further reduced
to transformations that do not interchange the Majorana spinors. But then 
also complex transformations are allowed, leading to the residual
$SU(N_\mathsf{f})_\mathsf{V}$ symmetry group. 

\begin{figure*}[tb]
\psset{xunit=1cm,yunit=1cm,runit=1cm}
\begin{center}
\begin{pspicture}(0,0.5)(12,4.5)
\rput(2,4){\rnode{A}{
\psframebox{$U(2N_\mathsf{f})$}}}
\rput(2,2){\rnode{B}{
\psframebox{$SU({2N_\mathsf{f}})
\otimes \mathbbm{Z}(2)_\mathsf{B}$}}}
\rput(10,2){\rnode{C}{
\psframebox{$SO(2 N_\mathsf{f})_\mathsf{V}\otimes \mathbbm{Z}(2)_\mathsf{B}$}}}
\rput(2,0){\rnode{D}{
\psframebox{$SU(N_\mathsf{f})_\mathsf{A}\otimes
SU(N_\mathsf{f})_\mathsf{V} \otimes
U(1)_\mathsf{B}/\mathbbm{Z}(N_\mathsf{f})$}}} 
\rput(10,0){\rnode{E}{
\psframebox{$SU(N_\mathsf{f})_\mathsf{V} \otimes
U(1)_\mathsf{B}/\mathbbm{Z}(N_\mathsf{f})$}}}
\ncline{->}{A}{B}\naput{anomaly}
\ncline{->}{B}{C}\naput{$m, \erw{\bar{\Psi} \Psi}$}
\ncline{->}{C}{E}\naput{$\mu$}
\ncline{->}{D}{E}\naput{$m,\erw{\bar{\Psi} \Psi}$}
\ncline{->}{B}{D}\naput{$\mu$}
\end{pspicture}
\end{center}
\vskip5mm
\caption{Pattern of chiral symmetry breaking in $G_2$-QCD.}
\label{fig:chiralSymmG2QCD}
\end{figure*}
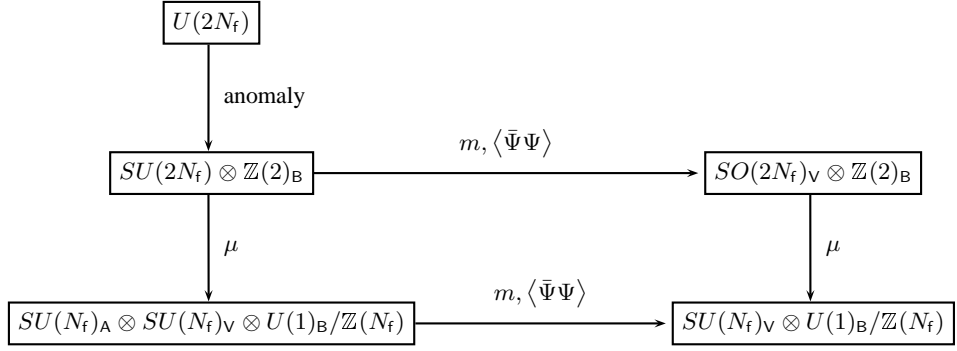

The pattern of chiral symmetry breaking in $G_2$-QCD is summarized in Figure \ref{fig:chiralSymmG2QCD}. If chiral symmetry is spontaneously broken, the axial chiral multiplet becomes
massless according to Goldstone's theorem. In contrast to QCD, because
of the extended chiral symmetry group, already in the case of a single
Dirac flavor it contains a
non-trivial $SU(2)$ and chiral symmetry breaking is
possible. This is one reason why in the following only $G_2$-QCD with a single
Dirac flavour $N_\mathsf{f}=1$ is investigated. The chiral symmetry is then given by
\begin{equation}
SU(2) \otimes \mathbbm{Z}(2)_\mathsf{B}.
\end{equation}
The corresponding creation operators for the Goldstone bosons are given by
\begin{equation}
\begin{aligned}
d(0^{++})=&\bar{\chi} \gamma_5 \eta = \bar{\Psi}^\chargec \gamma_5 \Psi-\bar{\Psi}\gamma_5\Psi^\chargec, \\
d(0^{+-})=&\frac{1}{\sqrt{2}}\left(\bar{\chi}\gamma_5\chi-\bar{\eta}\gamma_5\eta\right) = \bar{\Psi}^\chargec\gamma_5\Psi+\bar{\Psi}\gamma_5\Psi^\chargec.
\end{aligned}
\end{equation}
As usual, baryon number $n_\text{B}$ is here defined as the transformation behaviour of an 
operator under the $U(1)$ subgroup of the vector chiral transformation,
\begin{equation}
\Psi \mapsto e^{\ii n_\text{B} \alpha}\Psi,
\end{equation}
such that a quark has baryon number $n_\text{B}=1$ and an anti-quark $n_\text{B}=-1$.
With this definition of baryon number the Goldstone bosons have $n_\text{B}=2$. They are scalar diquarks instead of
pseudoscalar mesons as in QCD. 

\section{Spectroscopy for $N_\mathsf{f}=1$ $G_2$-QCD}\label{sspectheo}

\noindent The possible quark and gluon content of (colorless) bound states is determined by the tensor products of $G_2$-QCD. Quarks in $G_2$ transform under the $7$-dimensional 
fundamental representation, gluons under the $14$-dimensional fundamental
(and at the same time adjoint) representation. The decomposition of tensor 
products of the lowest-dimensional representations
into irreducible representations is given by
\begin{widetext}
\begin{equation}
\label{eq:representationsG2}
\begin{aligned}
\Rep{7} \otimes \Rep{7}&=\Rep{1} \oplus \Rep{7} \oplus \Rep{14} \oplus
\Rep{27},
\\
\Rep{7} \otimes \Rep{7} \otimes \Rep{7}&=\Rep{1} \oplus 4 \cdot \Rep{7} \oplus
2 \cdot \Rep{14} \oplus 3 \cdot \Rep{27} \oplus 2 \cdot \Rep{64} \oplus
\Rep{77'},
\\
\Rep{14} \otimes \Rep{14}&=\Rep{1} \oplus \Rep{14} \oplus \Rep{27} \oplus
\Rep{77} \oplus \Rep{77'}, \\
\Rep{14} \otimes \Rep{14}\otimes \Rep{14}&=\Rep{1}\oplus \Rep{7} \oplus 5\cdot
\Rep{14} \oplus 3\cdot\Rep{27}\oplus\dotsb, \\
\Rep{7} \otimes \Rep{14} \otimes \Rep{14} \otimes \Rep{14}&=\Rep{1} \oplus
\dotsb.
\end{aligned}
\end{equation}
\end{widetext}
\noindent
Thus we expect to find bound states for every integer quark number
$n_\text{q}$. Mesons have $n_\text{q}=0$, diquarks $n_\text{q}=2$, and nucleons $n_\text{q}=3$. 
In addition, there are more exotic bound states of gluons and quarks, for example a hybrid with $n_\text{q}=1$. 
Especially the latter state would be important, as the nucleons will only be stable in the chiral limit, if the hybrid is heavier than the nucleon. 
Of course, more complicated states with higher baryon numbers are possible, as well as glueballs, 
but are expected to play no role either in the vacuum, or at the moderate densities we investigate here.

In the following we give an overview over our implementation of possible bound states for $N_\mathsf{f}=2$, see Tables \ref{tab:nb0}-\ref{tab:nb3}. The subset of states 
of the $1$-flavour model, treated numerically below, are easily
recognized.

\begin{table}[htb]
\begin{tabular}{|c|c|c|c|c|c|}
\hline Name & $\mathcal{O}$ & $T$ & J & P & C\\
\hline $\pi$ & $\bar{u} \gamma_5 d$ & SASS & 0 & - & + \\
\hline $\eta$ & $\bar{u} \gamma_5 u$ & SASS & 0 & - & + \\
\hline $a$ & $\bar{u}d$ & SASS & 0 & + & + \\
\hline $f$ & $\bar{u}u$ & SASS & 0 & + & + \\
\hline $\rho$ & $\bar{u}\gamma_\mu d$ & SSSA & 1 & - & + \\
\hline $\omega$ & $\bar{u}\gamma_\mu u$ & SSSA & 1 & - & + \\
\hline $b$ & $\bar{u}\gamma_5 \gamma_\mu d$ & SSSA & 1 & + & + \\
\hline $h$ & $\bar{u}\gamma_5 \gamma_\mu u$ & SSSA & 1 & + & + \\
\hline
\end{tabular}\\
\caption{Bound states of $G_2$-QCD with $2$
flavours and baryon number $n_\text{B}=0$. For details see text.}
\label{tab:nb0}
\end{table}

\begin{table}[htb]
\begin{tabular}{|c|c|c|c|c|c|}
\hline Name & $\mathcal{O}$ & $T$ & J & P & C\\
\hline $N'$ & $T^{abc}(\bar{u}_{a} \gamma_5 d_{b}) u_{c}$ & SAAA & 1/2 & $\pm$ & $\pm$ \\
\hline $\Delta'$ & $T^{abc} (\bar{u}_{a} \gamma_\mu u_{b}) u_{c}$ & SSAS & 3/2 & $\pm$ & $\pm$ \\
\hline Hybrid & $\epsilon_{abcdefg} u^a F_{\mu\nu}^{bc} F_{\mu\nu}^{de} F_{\mu\nu}^{fg}$ & SSSS & 1/2 & $\pm$ & $\pm$ \\
\hline
\end{tabular}\\
\caption{Bound states with baryon number $n_\text{B}=1$. For details see text.}
\label{tab:nb1}
\end{table}

\begin{table}[htb]
\begin{tabular}{|c|c|c|c|c|c|}
\hline Name & $\mathcal{O}$ & $T$ & J & P & C\\
\hline $d(0^{++})$ & $\bar{u}^\chargec \gamma_5 u + c.c.$ & SASS & 0 & + & + \\
\hline $d(0^{+-})$ & $\bar{u}^\chargec \gamma_5 u - c.c.$ & SASS & 0 & + & - \\
\hline $d(0^{-+})$ & $\bar{u}^\chargec u + c.c.$ & SASS & 0 & - & + \\
\hline $d(0^{--})$ & $\bar{u}^\chargec u - c.c.$ & SASS & 0 & - & - \\
\hline $d(1^{++})$ & $\bar{u}^\chargec \gamma_\mu d - \bar{d}^\chargec \gamma_\mu u + c.c.$ & SSSA & 1 & + & + \\
\hline $d(1^{+-})$ & $\bar{u}^\chargec \gamma_\mu d - \bar{d}^\chargec \gamma_\mu u - c.c.$ & SSSA & 1 & + & - \\
\hline $d(1^{-+})$ & $\bar{u}^\chargec \gamma_5 \gamma_\mu d - \bar{d}^\chargec \gamma_5 \gamma_\mu u + c.c.$ & SSSA & 1 & - & + \\
\hline $d(1^{--})$ & $\bar{u}^\chargec \gamma_5 \gamma_\mu d - \bar{d}^\chargec \gamma_5 \gamma_\mu u - c.c.$ & SSSA & 1 & - & - \\
\hline
\end{tabular}\\
\caption{Bound states with baryon number $n_\text{B}=2$. For details see text.}
\label{tab:nb2}
\end{table}

\begin{table}[htb]
\begin{tabular}{|c|c|c|c|c|c|}
\hline Name & $\mathcal{O}$ & $T$ & J & P & C\\
\hline $N$ & $T^{abc}(\bar{u}_{a}^\chargec \gamma_5 d_{b}) u_{c}$ & SAAA & 1/2 & $\pm$ & $\pm$ \\
\hline $\Delta$ & $T^{abc} (\bar{u}_{a}^\chargec \gamma_\mu u_{b}) u_{c}$ & SSAS & 3/2 & $\pm$ & $\pm$ \\
\hline
\end{tabular}\\
\caption{Bound states with baryon number $n_\text{B}=3$. For details see text.}
\label{tab:nb3}
\end{table}

In all tables $\mathcal{O}$ is the interpolating operator used to extract the mass in simulations, $T$ the behaviour of the wave function under change of position, spin, colour and flavour (S stands for symmetric, A for anti-symmetric), and the spin ($J$), parity ($P$) and charge conjugation ($C$) quantum numbers. 
States with baryon number $0$ and $3$ are also present in QCD while the others are additional states of $G_2$-QCD.

In our simulations the states of the $2$-flavour model are included by partial quenching, that means we are dealing with two valence quark flavours, but only one sea quark flavour. 
In QCD, this is a surprisingly good approximation, see e.\ g.\ \cite{Engel:2010my}, and there is no obvious reason why this should be different in $G_2$-QCD.

There is one particular caveat, which is due to the limitation in computational resources for this project. The diquark correlation function that we measure on the lattice is given by
\begin{equation}
\begin{aligned}
C_{d}(x,y)=&\erw{d(0^{++})(x)\, d(0^{++})^\dagger(y)} \\ =&\erw{d(0^{+-})(x)\,
d(0^{+-})^\dagger(y)}\\=&\erw{ \contraction[2ex]{}{\bar{\chi}(x)}{\gamma_5
\chi(x)
\bar{\chi}(y) \gamma_5}{\chi(y)} 
\contraction[1ex]{\bar{\chi}(x) \gamma_5 }{\chi(x)}{}{\bar{\chi}(y)}
\bar{\chi}(x)\gamma_5 \chi(x)\, \bar{\chi}(y)\gamma_5,
\chi(y)}
\end{aligned}
\end{equation}
showing that the diquark masses are degenerate and its correlation functions
contain only connected contributions, like for example the correlation function for the pion in QCD.
The corresponding correlation function for the $\eta$ meson reads
\begin{equation}
\begin{aligned}
C_{\eta}(x,y)=&\erw{{\eta}(x)\, {\eta}^\dagger(y)}\\
=&2\erw{
\contraction[1ex]{}{\bar{\chi}(x)}{\gamma_5}{\chi(x)} 
\contraction[1ex]{\bar{\chi}(x) \gamma_5
\chi(x)}{\bar{\chi}(y)}{\gamma_5}{\chi(y))} \bar{\chi}(x)\gamma_5 \chi(x)\,
\bar{\chi}(y)\gamma_5 \chi(y)}+C_{d}(x,y)\label{etacorr}
\end{aligned}
\end{equation}
The difference between the $\eta$ and
the diquark correlation function is only the disconnected contribution. 
Therefore, uncertainties in the treatment of the disconnected contribution can blur the line between the $\eta$ and the diquarks.

Analog relations lead for the partially quenched calculations performed here to some relations between flavour singlet diquark masses and flavour non-singlet meson masses,
\begin{equation}
\begin{aligned}
 m_{d(0^+)}=&m_{\pi(0^-)}\\
m_{d(0^-)}=&m_{a(0^+)}\\
m_{d(1^+)}=&m_{\rho(1^-)}\\
m_{d(1^-)}=&m_{b(1^+)}.
\end{aligned}
\end{equation} 
Thus, for every diquark there is a flavour non-singlet meson with the same mass but opposite parity.

\section{Algorithmic considerations}\label{stech}

\noindent
In our lattice simulations we use a Hybrid Monte-Carlo algorithm \cite{Duane:1987de} to 
generate the probability distribution. Our implementation is based on \cite{Wellegehausen:2011sc}, where the algorithm was applied to $G_2$-Yang-Mills-Higgs theory. 

For the gauge action we choose the tree-level improved Symanzik gauge action
\cite{K1983187,K1983205,Luscher:1984xn,PhysRevD.78.054504}
\begin{equation}
\begin{aligned}
S[\gU]=\frac{\beta}{N_\mathsf{c}}\left \lbrace c_0 \sum
\limits_{\square} \tr \left(1-\Re \group{U}_\square\right)+ \right. \\ \left. c_1
\sum \limits_{\square\square} \tr \left(1-\Re
\group{U}_{\square\square}\right)\right \rbrace.
\end{aligned}
\label{eq:generalGaugeAction}
\end{equation}
Here, $\group{U}_{\square}$ stands for the plaquette variable and
$\group{U}_{\square\square}$ for a rectangular path around two plaquettes. The parameters are given by
$c_0=1-8 c_1$, $c_1=-1/2$. Note that our convention is to factorize the number of colors from $\beta$.

For the fermion part, we use the ordinary Wilson action without improvements \cite{Gattringer:2010zz}. 
Though we cannot expect good chiral properties in this case, we can avoid rooting for staggered fermion. 
Using unrooted staggered fermions, and thus four flavours, would on the one hand create far too many Goldstone bosons, and would possibly put the theory too close or in the conformal window, according to the two-loop $\beta$-function. 
Fermion implementations with better chiral properties are unfortunately beyond our numerical resources.

For the fermion determinant we use pseudo-fermions together with a 
rational approximation of the inverse fermion matrix (RHMC algorithm) \cite{Kennedy:1998cu}. In the case of Dirac fermions the path
integral is given by \footnote{Below, $\tr$ denotes the integral over
$d$-dimensional space-time and the trace over all internal degrees of freedom.}
\begin{equation}
\begin{aligned}
\mathcal{Z}=&\int \mathcal{D}\Psi
\mathcal{D}\bar{\Psi}\mathcal{D}\group{U}e^{-S[\group{U}]-\tr\bar{\Psi}D\Psi}\\
=&\mathcal{N}\int \mathcal{D}\group{U}
\det\left(D[\group{U}]\right) e^{-S[\group{U}]}\\
=&\mathcal{N}\int \mathcal{D}\group{U}
\det\left(M[\group{U}]^
\frac{1}{2}\right) e^{-S[\group{U}]},
\end{aligned}
\end{equation}
where $D$ is the fermion operator and $M=D^\dagger D$ is a Hermitian and positive operator. Introducing $N_\mathsf{PF}$
complex-valued pseudo-fermions $\phi$ \cite{Weingarten:1980hx}, one can write
the partition function as
\begin{equation}
\begin{aligned}
\mathcal{Z}=\int \mathcal{D}\group{U}\mathcal{D}\phi
\exp\{-S_\mathsf{B}[\group{U},\phi]\}\quad \text{with} \\
S_\mathsf{B}[\group{U},\phi]=S[\group{U}]+\tr\,\sum
\limits_{p=1}^{N_\mathsf{PF}} \phi_p^\dagger M^{-q}\phi_p,
\end{aligned}
\end{equation}
where $S_\mathsf{B}$ is the bosonic action and $q$ is given by $q=\frac{1}{2
N_\mathsf{PF}}$. In the RHMC dynamics $M^{-q}$ is replaced by a rational
approximation according to
\begin{equation}
r(x)=x^{-q}\approx\alpha_0+\sum
\limits_{r=1}^{N_\mathsf{R}}\frac{\alpha_r}{x+\beta_r}.
\label{eq:rapproximation}
\end{equation}
For any rational number $q$ the coefficients $\alpha$ and $\beta$ can be
calculated with the Remez algorithm \cite{Fraser:1965:SMC:321281.321282}. The
numerical accuracy of the approximation in the interval $
I=[x_\mathsf{Min},x_\mathsf{Max}]$ depends on the number of terms $N_\mathsf{R}$ 
in (\ref{eq:rapproximation})
and the numerical accuracy of the coefficients $\alpha$ and $\beta$. In the following
$r_\mathsf{S}(x),\,\mathsf{S}=\{I,\epsilon,q\}$ denotes a rational approximation
of the function $x^{-q}$ with $\epsilon=\sup \limits_{x \in I}\norm{r(x)-x^{-q}}$.

In order to obtain an exact update algorithm, the bosonic action is written in
the form
\begin{equation}
\begin{aligned}
S_\mathsf{B}[\group{U},\phi]=
S[\group{U}]+S_{\mathsf{md}}(M)+S_{\mathsf{acc}}(M)+S_{\mathsf{rw}}(M),
\end{aligned}
\end{equation}
where the different contributions are given by
\begin{equation}
\begin{aligned}
S_{\mathsf{md}}=&\tr\,\sum \limits_{p=1}^{N_\mathsf{PF}} \phi_p^\dagger
r_{\mathsf{S}_\mathsf{md}}\phi_p,\\
S_{\mathsf{acc}}=&\tr\,\sum \limits_{p=1}^{N_\mathsf{PF}} \phi_p^\dagger
\left(r_{\mathsf{S}_\mathsf{acc}}(M)-r_{\mathsf{S}_\mathsf{md}}(M)\right)\phi_p,\\
S_{\mathsf{rw}}=&\tr\,\sum\limits_{p=1}^{N_\mathsf{PF}}
\phi_p^\dagger \left(M^{-q}-r_{\mathsf{S}_\mathsf{acc}}(M)\right)\phi_p\,.
\end{aligned}
\end{equation}
The sum $S[\group{U}]+S_{\mathsf{md}}(M)$ is used in the calculation of the HMC
molecular dynamics, the sum
$S[\group{U}]+S_{\mathsf{md}}(M)+S_{\mathsf{acc}}(M)$ in the Metropolis acceptance 
step of the HMC algorithm and the last term $S_{\mathsf{rw}}(M)$ in a
reweighting step to assure an exact update algorithm.

In practice, the reweighting step is not necessary since it is more efficient to
choose $r_{\mathsf{S}_\mathsf{acc}}$ such that it approximates $M^{-q}$ up to machine precision. For
the generation of the pseudo-fermion fields from a Gaussian distributed
vector the square root of $M^{q}$ is needed as well. This is achieved by an
approximation $r_{\mathsf{S}_\mathsf{pf}}(M)\approx M^{q/2}$. To obtain an exact update algorithm, the following choices are made,
\begin{equation}
\begin{aligned}
r_{\mathsf{S}_\mathsf{pf}}(M)=&\{I \supseteq \Sigma(M),10^{-16},-q/2\},\\r_{\mathsf{S}_\mathsf{acc}}(M)=&\{I \supseteq \Sigma(M),10^{-16},q\},
\end{aligned}
\end{equation}
where $\Sigma(M)=[\lambda_{\mathsf{min}},\lambda_{\mathsf{max}}]$ is the
spectral range of the Hermitian operator $M$. In most of the simulations, an
approximation for the pseudo-fermion and acceptance step approximation of degree
$N_\mathsf{R}=25$ is used in an interval $I=[10^{-7},10]$.

The free parameters left to optimize the algorithm are the integration scheme used in the molecular dynamics
and the degree and approximation range of the molecular dynamics rational
approximation $r_{\mathsf{S}_\mathsf{md}}(M)$. The inversions of the matrix $M$ in the
rational approximations are calculated with a multiple-mass conjugate gradient
solver (MMCG) \cite{Jegerlehner:1996pm} which is able to compute all terms of
\eqref{eq:rapproximation} within a single inversion of the fermion matrix $M$.

\subsection{Symplectic integration and multiple time scales}

\noindent
In order to speed up our simulation, we use integration on different time scales in an HMC trajectory. 
The simplest possible integration scheme is the leap-frog
scheme \cite{4332919}. The time evolution $T$ from $\tau=0$ to $\tau=t_\mathsf{HMC}$ with step size $\delta \tau=\frac{t_\mathsf{HMC}}{n}$ with the leap-frog
time evolution operator $T_\mathsf{LF}$ can be written as
\begin{equation}
\begin{aligned}
T(t_\mathsf{HMC},\delta \tau)= &T_\mathsf{LF}(\delta \tau)^{n}\,, \\
T_\mathsf{LF}(\delta \tau)= &T_S(\frac{1}{2}\delta \tau)\,T_\gU(\delta \tau)\,T_S(\frac{1}{2}\delta
\tau)\,,
\end{aligned}
\end{equation}
where $T_S$ describes time evolution for the momenta and $T_\gU$ for the fields.
An improved second-order integrator is
given by the Sexton-Weingarten scheme \cite{Sexton1992665},
\begin{equation}
\begin{aligned}
T_\mathsf{SW}(\delta \tau)=&T_S(\frac{\delta \tau}{6})\, T_\gU(\frac{\delta
\tau}{2}) \\ \times &T_S(\frac{2 \delta \tau}{3})\, T_\gU(\frac{\delta
\tau}{2})\,T_S(\frac{\delta \tau}{6}).
\end{aligned}
\end{equation}
A fourth order integrator is given by \cite{PhysRevE.73.036706}
\begin{equation}
\begin{aligned}
T_\mathsf{4}(\delta \tau)=T_S(\rho \delta \tau)\,T_\gU(\lambda \delta
\tau)\,T_S(\theta \delta
\tau)\\ \times  T_\gU((1-2\lambda)\frac{\delta
\tau}{2})\,T_S((1-2(\theta+\rho)) \delta \tau)\\
\times  T_\gU((1-2 \lambda)\frac{\delta
\tau}{2})\,T_S(\theta \delta \tau)\\ \times T_\gU(\lambda \delta
\tau)\,T_S(\rho\delta \tau),
\end{aligned}
\end{equation}
with parameter values 
\begin{equation}
\begin{aligned}
\rho=&0.1786178958448091, \\
\theta=&−0.06626458266981843 \quad \text{and}\\
\lambda=&0.7123418310626056.
\end{aligned}
\end{equation}
Higher order integrators are constructed in
\cite{Haruo1990262}. Further improvement can be achieved by integration on
multiple time scales \cite{Jansen:2005yp}. For this purpose an
arbitrary integrator $T_\mathsf{s}$ (here s stands for the integration scheme)
is written as a function of the basic time evolution operators $T_S$
and $T_\gU$ and the integration step size
$\delta \tau$, $T_\mathsf{s}=T_\mathsf{s}(T_S,T_\gU,\delta
\tau)$.

If the action can be written as a sum of contributions $S_j$, i.e.
$S=S_1+S_2+\dots$, then multiple time scale integration can be defined by the recursion relation
\begin{equation}
\begin{aligned}
T_{\mathsf{s}_j}^j (T_{S_j},T_\gU,\delta
\tau_j)=\\T_{\mathsf{s}_j}^j
(T_{S_j},[\,T_{\mathsf{s}_{j-1}}^{j-1}(T_{S_{j-1}},T_\gU,\delta
\tau_{j}/n_j)\,]^{n_j},\delta \tau_j)\,,
\end{aligned}
\end{equation}
where $S_j$ denotes the subset of the action that should be taken into
account in the computation of the `force' on the $j$-th time scale with
step size $\delta \tau_j$. Here, we often use a two time-scale integration, which is a combination of the Sexton-Weingarten scheme with the leap-frog scheme,
\begin{equation}
\begin{aligned}
T(\delta \tau)=&T_{S_0}(\frac{\delta
\tau}{2})\,T_\mathsf{SW}(T_{S_1},T_\gU,\delta
\tau)\,T_{S_0}(\frac{\delta \tau}{2})\\
=&T_{S_0}(\frac{\delta
\tau}{2})\,T_{S_1}(\frac{\delta \tau}{6}) \, T_\gU(\frac{\delta
\tau}{2})\\ \times &\,T_{S_1}(\frac{2 \delta \tau}{3})\, T_\gU(\frac{\delta
\tau}{2})\,T_{S_1}(\frac{\delta
\tau}{6})\,T_{S_0}(\frac{\delta \tau}{2}).
\end{aligned}
\end{equation}
Here, the `force' according to $S_1$ has to be calculated twice as
often as the `force' belonging to $S_0$.

Another scheme often used
is the combination of a fourth order integrator with the
Sexton-Weingarten scheme or with the simple leap-frog scheme.
Multiple-time-scale integration is efficient if parts of the action with large
contribution to the HMC `force' are cheap in computation time.

\subsection{Optimization of the RHMC algorithm}

\noindent
The efficiency of the RHMC algorithm depends crucially on the lowest
eigenvalues entering the condition number $\kappa \approx
\lambda_\text{max}/\lambda_\text{min}$ of the Hermitian operator used in the
rational approximation. The number of total inversion steps for a given precision $\delta_{\max}$
(the inversion precision for the lowest mass, i.e. the lowest value of
$\beta_r$) in the MMCG solver increases significantly with decreasing values of the constants $\beta_r$ in the rational approximation. 
Fortunately, the force contribution in the RHMC algorithm is for small constants
also significantly lower than for larger constants (the reason is that $\alpha_r$ decreases also
with decreasing $\beta_r$). Only in the case of very small eigenvalues, the
force from these lowest eigenmodes becomes more important.

This feature of the RHMC algorithm can now be used to
optimize the algorithm with respect to computation time. Two different
strategies are useful: The first is to integrate the terms with smaller
$\beta_r$ on a coarser time scale than the terms with larger $\beta_r$, i.e.
larger force. The second is to increase the lower bound of the
approximation interval, resulting in larger values of $\beta_r$ and a possibly
smaller degree of the rational function used for the molecular dynamics. This
reduces the number of CG-steps for a given inversion precision $\delta_{\max}$
significantly.

Further optimization can be achieved by increasing the
precision $\delta_{\max}$ used for the inversion, leading also to a significantly reduced number of
CG-steps. The best choice of course depends on the given problem and is in
general a combination of both strategies. Further optimizations implemented include
even-odd preconditioning \cite{Allton1993331} as well as an exact computation of a few
lowest eigenvalues in the MMCG solver. According to \cite{Clark:2004cq}, the
optimal number of pseudo-fermions is roughly given by the condition number of
the fermion matrix, $N_\mathsf{PF}^{\text{opt}}\approx \frac{1}{n}\ln
\kappa(M)$.

\subsection{Fermionic correlation functions}

\noindent
For the computation of the connected part of the correlation function, the
fermion matrix is inverted on a point-like source in space and time at a
randomly chosen lattice point $y$, leading to the point-to-all propagator.
Here, $N_\mathsf{c} \times N_\mathsf{s}$ (number of colours times the dimension of
the representation of the Clifford algebra) inversions of the fermion matrix
with the CG solver have to be made.

The disconnected diagrams, and for instance
observables like the chiral condensate or the quark number density, are calculated with the stochastic estimator technique (SET) \cite{Dong:1993pk,Bitar1989348}. 
Here every element of the fermion propagator is calculated as an ensemble
average over a noisy estimator $\eta$,
\begin{equation}
\begin{aligned}
\tilde{\Delta}_{ij}=\lim\limits_{N_\text{est} \rightarrow
\infty}\erw{\eta_j^\dagger \chi_i} \quad \text{with} \\
\chi=\tilde{\Delta}\eta \quad\text{and}\quad
\lim\limits_{N_\text{est} \rightarrow
\infty}\erw{\eta^\dagger_i \eta_j}=\delta_{ij}.
\label{eq:SEST}
\end{aligned}
\end{equation}
In practice, the ensemble average is taken over a finite number of
$N_\text{est}$ noisy estimators, where the source $\eta$ is given by
Gaussian or $\mathbbm{Z}(2)$ noise, satisfying the last equation in
\eqref{eq:SEST}. The sink is again calculated with a CG solver, making a total of
$N_\text{est}$ matrix inversions to obtain an estimator for every matrix element
of the propagator. In the case of local lattice averaged observables, like the chiral
condensate, a number of $N_\text{est}\approx 10$ estimators is sufficient to
get a reliable result. For the disconnected part of four-point correlation
functions (many) more estimators are necessary.

We note that we extract masses from the correlators $C(t)$ by fits of the type
\begin{equation}\label{eq:mandmstar}
C(t)=a\cosh(m t)+b\cosh(m^* t),
\end{equation}
\noindent or with a single $\cosh$-fit, where a double-$\cosh$ fit was not possible. 
The quoted errors denote only the statistical error from a simultaneous
up- or down-shift of the correlation function by one standard deviation.

We identify the smaller of the two parameters $m$ and $m^*$ in
(\ref{eq:mandmstar})
as the ground state mass, and mark the next higher mass with an asterisk '*'. 
We do not make any attempt to identify whether these are genuine excited states or merely scattering states, and, as noted in section \ref{sspectheo}, we use a single operator per quantum number channel. 
We also do not attempt to identify whether the lowest state is a genuine bound state or a scattering state, even if it appears energetically favorable for them to decay. 
For some states we are also limited by statistics, and thus could not measure the mass of all relevant channels. 
This applies especially to the hybrids. We therefore have to assume in the following that at least the ground states are reasonably stable states.

\section{Lattice spectroscopy results}\label{sspectrum}

\noindent
In order to fix our parameters we compute the diquark masses and the proton mass for different parameters
of the inverse gauge coupling $\beta$ and the hopping parameter $\kappa$ on a $8^3 \times 16$ lattice. 
We make here the implicit assumption that the nucleon is (quasi-) stable, i.\ e.\ it is not energetically favorable or possible for it to decay into a hybrid and a diquark. 
Since the hybrids were too noisy to obtain reliable results, we could not check this assumption.

\begin{figure}[htb]
\scalebox{1.0}{\input{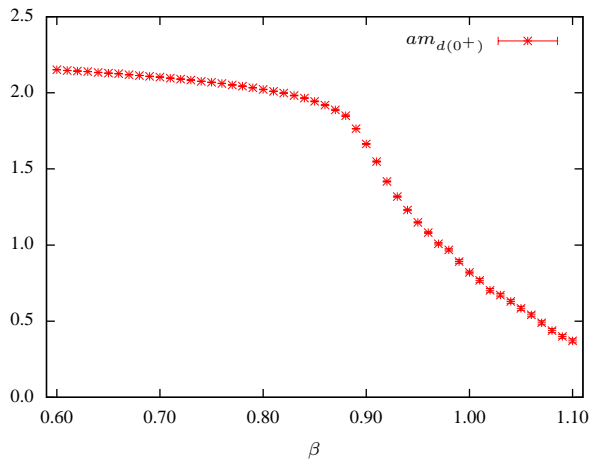}}
\caption{Mass of the pseudo Goldstone boson as a function of $\beta$ for $\kappa=0.147$.}
\label{fig:massDiquark}
\end{figure}

\begin{figure}[htb]
\scalebox{1.0}{\input{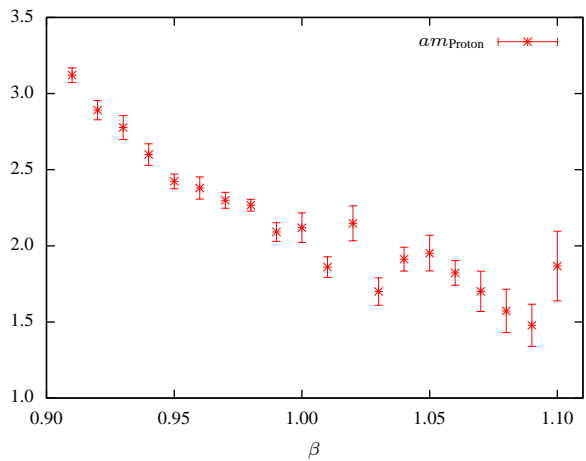}}
\caption{Mass of the proton as a function of $\beta$ for $\kappa=0.147$.}
\label{fig:massProton}
\end{figure}

To assess the distance from the chiral limit, we first compare the Goldstone sector to the nucleon sector. In Figure \ref{fig:massDiquark} the $d(0^{+})$ mass is shown as a function of the 
inverse gauge coupling $\beta$ for a fixed value of the hopping parameter $\kappa$. In Figure \ref{fig:massProton} the proton mass is plotted for the same parameters.

Care has to be taken, as $G_2$-QCD possesses an unphysical lattice bulk phase at strong coupling where monopoles condense. 
The critical inverse gauge coupling for the transition to the physical weak coupling phase depends on the hopping parameter. 
For $\kappa=0.147$ it is located around $\beta\approx 0.90$. 
We observe that in the bulk phase the lattice diquark mass is only weakly dependent on the gauge coupling and therefore the lattice spacing does not depend on $\beta$. 
Above the transition, the lattice diquark mass decreases with increasing inverse gauge coupling. 
Since the bulk transition is a crossover (at least for infinitely heavy quarks \cite{Pepe:2006er,Cossu:2007dk}) we have to choose a gauge coupling for our simulations that is far above the transition point. 
For our spectroscopy results we have checked that the monopole density is always below one percent of the monopole saturation density in the bulk phase.

\begin{figure}[htb]
\scalebox{1.0}{\input{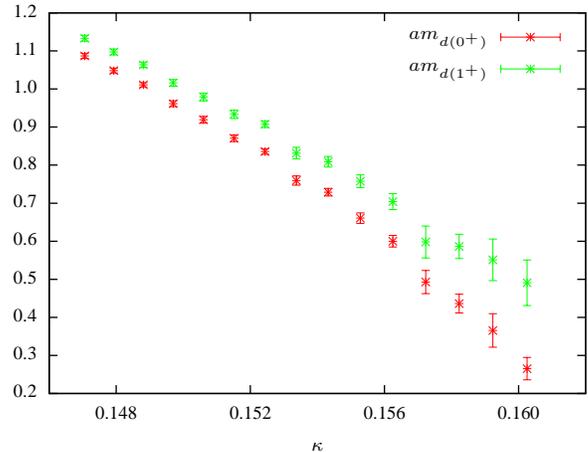}}
\caption{Mass of the $0^{+}$ and the $1^{+}$ diquark as a function of $\kappa$ for $\beta=0.96$.}
\label{fig:massDiquarks}
\end{figure}

\begin{table*}[htb]
\begin{tabular}{|c|c|c|c|c|c|c|c|c|}
\hline  Ensemble & $\beta$ & $\kappa$ & $m_{d(0^+)} a$ & $m_N a$  & $m_{d(0^+)}$ [MeV] & $a$ [fm] & $a^{-1}$ [MeV]& MC\\
\hline Heavy & $1.05$ & $0.147$ & $0.59(2)$ & $1.70(9)$ & $326$ & 0.357(33) & 552(50) & 7K   \\
\hline Light & $0.96$ & $0.159$ & $0.43(2)$ & $1.63(13)$ & $247$ & 0.343(45) & 575(75) & 5K \\
\hline
\end{tabular}\\
\caption{Parameters for two different ensembles. All results are from a $8^3\times 16$ lattice.}
\label{tab:ensembles}
\end{table*}

For heavy quarks the ratio of diquark and proton mass should be 2/3 while it should go to zero in the chiral limit. A second mass ratio to fix the bare parameters is the ratio
of the $0^{+}$ and the $1^{+}$ diquark. For heavy quarks only the number of quarks is important and the ratio should be one while in the chiral limit the 
spin zero diquark becomes massless while the spin one diquarks stay massive. The results for the masses are shown in Figure \ref{fig:massDiquarks} as a function of $\kappa$ and fixed $\beta$.
Indeed we see that for smaller Goldstone masses the ratio increases. In the following we discuss two different ensembles with parameters shown in Table \ref{tab:ensembles}. 
In the following, we will set our mass scale by the proton mass, $m_N=938$ MeV.

\begin{figure}[htb]
\scalebox{0.9}{\input{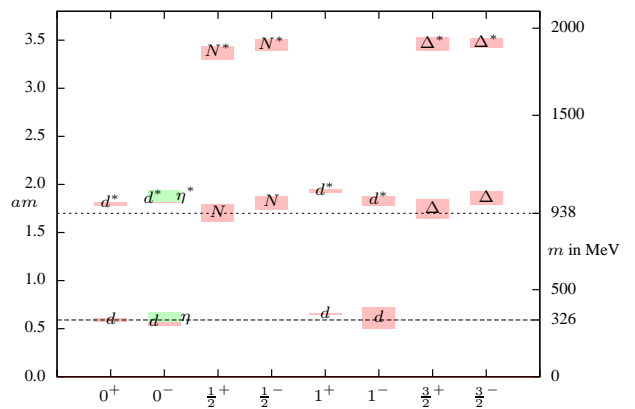}}
\caption{Mass spectrum of the heavy ensemble}
\label{fig:massHeavy}
\end{figure}

The mass spectrum for the heavy quark ensemble is shown in Fig.~\ref{fig:massHeavy}. The diquark masses are almost degenerate. Also the $\eta$ has essentially the same mass as the diquarks. 
For the nucleons there is almost no mass splitting between parity even and odd states.

\begin{figure}[htb]
\scalebox{0.9}{\input{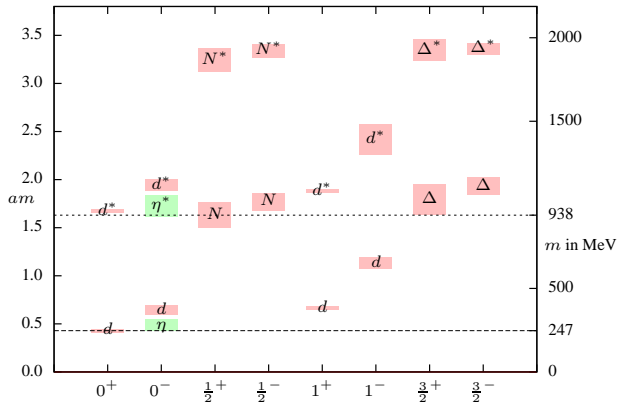}}
\caption{Mass spectrum of the light ensemble}
\label{fig:massLight}
\end{figure}

In the light ensemble, shown in Figure \ref{fig:massLight}, the diquark masses are no longer degenerate. 
We observe a significant mass splitting between parity even and odd states as well as between scalar and vector diquarks. 
Especially, the Goldstone boson becomes the lightest state, with the $\eta$ also being somewhat heavier. 
This mass difference comes entirely from the disconnected part of the meson correlation function in (\ref{etacorr}). 
For the nucleons we also observe different masses for parity even and odd states and the spin 1/2 and spin 3/2 representations. 
Thus, the spectrum is indeed consistent with spontaneous chiral symmetry breaking, in accordance with quenched \cite{Danzer:2008bk} and previous results \cite{Maas:2012wr}. 
Especially, we find three clearly different scales in the light spectrum: A Goldstone scale, an intermediate boson scale set by the remaining diquarks, and the nucleon scale set by the $N$ and $\Delta$.

\section{$G_2$-QCD at zero temperature and finite baryon density}\label{sdensity}

\subsection{Scales at finite density}

\noindent
In \cite{Maas:2012wr} we already provided an overview over the full phase diagram of $G_2$-QCD as a function of temperature and baryon density. 
We will now show that the different hadronic scales observed in the spectra in Figs. \ref{fig:massHeavy} and  \ref{fig:massLight} reflect themselves in the structure of the finite density phase diagram.

The first scale, the Goldstone scale, must be related to the onset transition to baryonic matter, since the Goldstones carry quark number. 
This follows immediately from the silver blaze property of quantum field theories \cite{Cohen:2003kd} at zero temperature and finite density.

\begin{figure}[htb]
\scalebox{1.0}{\input{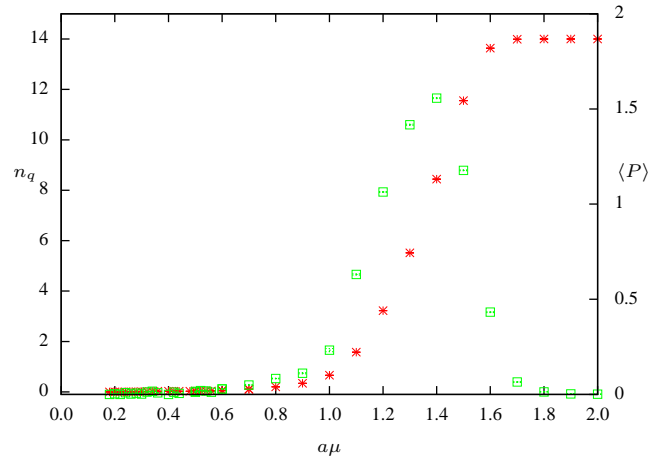}}
\caption{The quark number density (red) and the Polyakov loop (green) as a function of chemical potential are shown.}
\label{fig:saturation}
\end{figure}

To investigate this regime, we have calculated the quark number density $n_q$ given by
\begin{equation}
n_q=\frac{1}{V} \frac{\partial\ln Z}{\partial\mu}.
\end{equation}
In \cite{Maas:2012wr} we observed that for small values of the chemical potential the system
remains in the vacuum, i.\ e.\ the quark number density vanishes, which is expected
due to the silver blaze property. When increasing the chemical
potential further the quark number density starts rising, indicating that baryonic matter is
present and the system is no longer in the vacuum state. At even larger values of $\mu$ the quark number density
saturates. The value of the saturation matches the theoretical prediction of $n_{q,\text{max}}=2 N_\mathsf{c}=14$ \cite{Maas:2012wr}. This is depicted in
Fig. \ref{fig:saturation}. 

The same figure shows the dependence of the Polyakov loop 
on the chemical potential from $\mu=0$ up to saturation. The decrease of the Polyakov loop close to saturation also indicates that the system enters a quasi-quenched phase, where the quark dynamics freezes out \cite{Hands:2006ve,Maas:2012wr}. 
This emphasizes that for $a\mu\approx 1$ lattice artifacts start to dominate the system. 
However, this is for both ensembles at an already high quark chemical potential of about 550 MeV, corresponding to a nucleon chemical potential of 1.65 GeV.

\begin{figure}[htb]
\scalebox{1.0}{\input{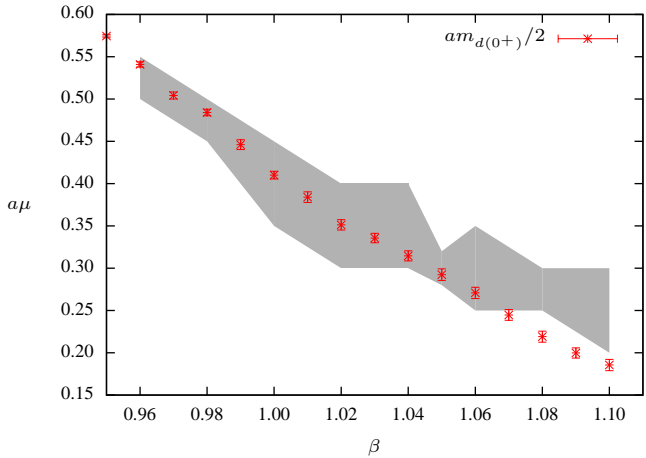}}
\caption{The onset transition observed in the quark number density is compared to half of the mass of the lightest state, the $0^+$ diquark, for different gauge couplings $\beta$, and thus different quark masses.}
\label{fig:onset}
\end{figure}

A closer look into this phase diagram at zero temperature shows that the quark number density already jumps, or very quickly rises, to a very small but nonzero value already at a very small chemical potential. In Figure \ref{fig:onset}
this onset transition is compared to half of the mass of the lightest baryon, the Goldstone $0^+$ diquark. For various values of $\beta$ very good agreement is found.
This is the expected manifestation of the silver blaze property for baryon chemical potential, i.\ e.\ half of the
mass of the lightest bound state carrying baryon number is a lower bound for the onset transition to a non-vacuum state\footnote{Note that a finite lattice is strictly speaking never at zero temperature, and therefore the silver blaze property is never exactly realized. 
However, such violations are expected to be exponentially suppressed by the spatial volume, which effectively determines the residual temperature. 
We do indeed observe such artifacts.}.

\begin{figure}[htb]
\scalebox{1.0}{\input{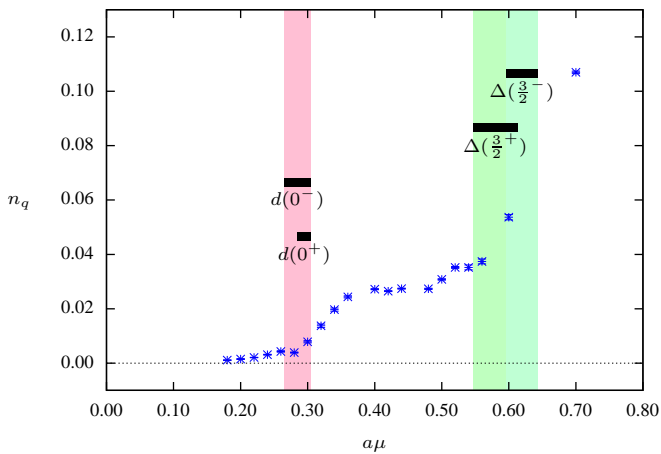}}
\caption{Shown is the quark number density compared to baryon mass divided by baryon number for the \emph{heavy ensemble}.}
\label{fig:onset105}
\end{figure}

\begin{figure}[htb]
\scalebox{1.0}{\input{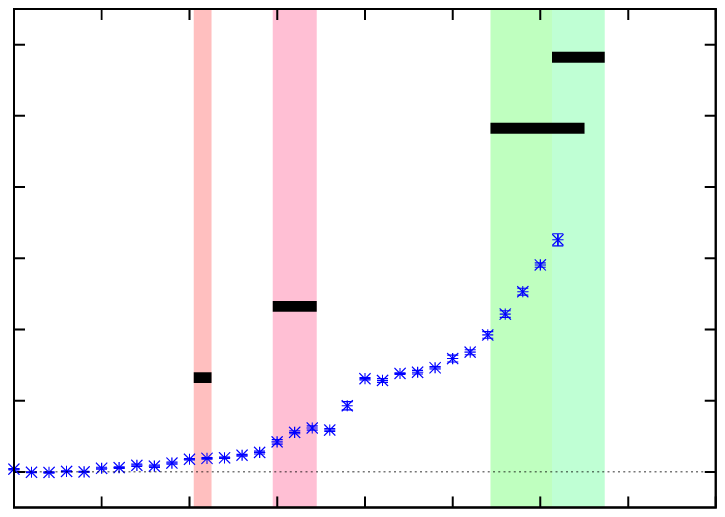}}
\caption{Shown is the quark number density compared to baryon mass divided by baryon number for the \emph{light ensemble}.}
\label{fig:onset096}
\end{figure}

For larger values of the chemical potential a series of plateaus develop where the quark
number density is almost constant, see Figure \ref{fig:onset105} for the heavy ensemble and Figure \ref{fig:onset096} for the light ensemble. 
In both cases, we observe at intermediate chemical potential interesting structures, which will be discussed below. 
At around $a\mu=0.6$ for the heavy ensemble and $a\mu=0.55$ for the light ensemble the quark number density starts increasing again and no further plateau is observed. 

It is quite interesting to compare these transitions to the masses of the diquarks and baryons normalized by their baryon number.

For the heavy ensemble, in addition to the silver blaze transition due to the diquark states we find good agreement of the $\Delta$ mass with the point where the quark number density increases without building a plateau.

For the light ensemble the two transitions at $a\mu\approx 0.22$ and $a\mu \approx 0.32$, each followed by a plateau, see Figure \ref{fig:onset096}, can be related to the observation of the splitting of the $0^+$ and $0^-$ diquark masses. 
Again the transition at $a\mu \approx 0.55$ is in good agreement with the $\Delta$ mass divided by three.

For both ensembles our observation is thus that transitions in the quark number density coincide with hadron masses divided by their baryon number. 
For a bosonic hadron a plateau is formed after the transition while for a fermionic hadron the quark number density increases further with increasing chemical potential. 
In both ensembles we observe also a transition at $a\mu\approx0.52$ (heavy ensemble) and $a\mu\approx0.38$ (light ensemble) that does not coincide with any of our spectroscopic states. 
Since this transition is followed by a plateau we speculate that this state might also be a bosonic hadron. 
A possible candidate could for example be a bound state of four quarks. However, this may also relate to some of the known states, if their masses turn out to be significantly dependent on the chemical potential. 
It is also possible that additional collective excitations arise, if any of the phases sustain a Bose-Einstein condensate, as has been argued for the low-density phase in two-color QCD \cite{Kogut:2000ek,Hands:2000ei,Hands:2006ve,Hands:2011ye,Strodthoff:2011tz,Strodthoff:2013cua,Boz:2013rca}.

This question is not simple to decide, as it is not clear how to reliably and unambiguously determine the mass of (quasi-) particles at finite density in lattice simulations. However, it will be crucial to understand it in the future.

\subsection{Free fermions}

\noindent Further interesting insights can be gained by comparing the results 
with the corresponding ones for non-interacting systems of fermionic particles. 
On the one hand, this can test whether the idea of (quasi-free) fermions or fermionic quasi-particles describe the theory adequately at some densities. 
On the other hand, the saturation effects should also yield a quasi-free behavior, indicating the onset of
lattice artifacts. We will only consider here the heavy ensemble, as for the light ensemble the acceptance rate dropped seriously in the range of $a\mu=0.7$ to $a\mu\approx 1.5$, and we can therefore not really assess the intermediate and saturation regime yet.

\begin{figure}[htb]
\scalebox{1.0}{\input{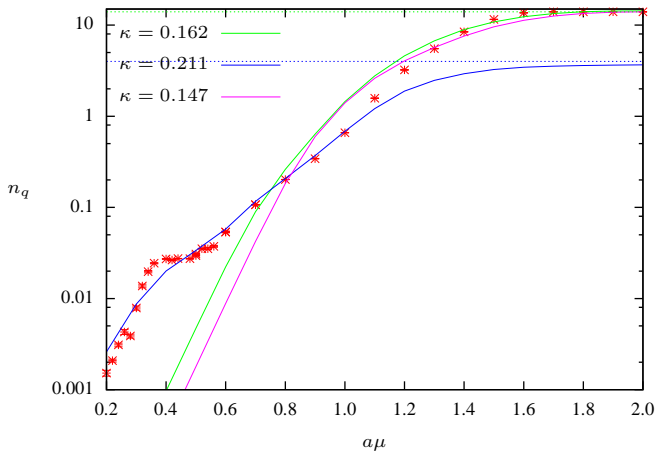}}
\caption{Fit of the quark number density for the heavy ensemble with the density for free lattice fermions.}
\label{fig:fit_105}
\end{figure}

We begin with the fermion density for a theory of free lattice (Wilson)
fermions with mass $\tilde{m}=m+d=1/(2\kappa)$. It can be derived in analogy to the staggered result of \cite{Hands:2000ei} and is given by
\begin{equation}
\begin{aligned}
n_\mathsf{f}^\text{free}(\mu,\tilde{m})/ n_\mathsf{f}^\text{sat}=&\\
\sum_{\tilde{p}} \frac{2 \ii \,\sqrt{1-\tilde{p}_0^2}\left(\sum_i \tilde{p}_i-\tilde{m}\right)}{4+\tilde{m}^2-2\tilde{m}\sum_\mu \tilde{p}_\mu\left(\sum_{\nu>\mu}\tilde{p}_\nu-\tilde{m}\right)}
\end{aligned}
\end{equation}
where the sum extends over all lattice momenta 
\begin{equation}
\begin{aligned}
\tilde{p}_0&=\cos\left(\frac{2\pi}{N_t}\left(k_0+\frac{1}{2}\right)-\ii \mu\right) \quad \text{and} \\
\tilde{p}_i&=\cos\left(\frac{2\pi k_i}{N_s}\right) \quad \text{with} \\ 
k_0&=1 \dots N_t \quad \text{and}\quad k_i=1 \dots N_s .
\end{aligned}
\end{equation}
When we tried to fit our data for the heavy ensemble to this formula with fitting parameters $\kappa$ (which enters $\tilde m$) 
and $n_\mathsf{f}^\text{sat}$ we observed that the behaviour changes at around $a \mu \approx 1$, see Figure \ref{fig:fit_105}. 
Above $a \mu=1$ the best fit for the data yields $\kappa=0.162$ and $n_\mathsf{f}^\text{sat}=14.4$. This is in good agreement with the values for free quarks of $\kappa=0.147$ and $n_\mathsf{f}^\text{sat}=n_q^\text{sat}=14$.
Although we expect that for very large values of $\mu$ the theory is exactly described by free quarks, in this intermediate region the Polyakov loop is not constant, and also the contribution of gluons to the free energy has not yet reached its quenched limit \cite{Maas:2012wr}. 
This might explain deviations from the exact values.
Still, the rather good fit suggests strongly that for $a \mu>1$ lattice artifacts become important.

Below $a \mu=1$ the data are very good described by $\kappa=0.211$ and $n_\mathsf{f}^\text{sat}=4.02$. 
The theoretical value for the saturation of a lattice gas of free $\Delta$-baryons is $n_B^\text{sat}=4$. This suggests  that between
$a \mu\approx 0.6$ and $a \mu\approx 1.0$ the main contribution to the quark number density may come from fermionic baryons, in agreement with our findings in the last section. 
Somewhat surprisingly these fermionic baryons would behave very much like a non-interacting gas. One should note, however, that formally the $\kappa$ value yields a negative mass. 
This is a consequence of using Wilson fermions. In principle we would have to correct for the additive mass shift. 
However, we do not yet know $\kappa_\text{critical}$ to do so. Determining it will require substantial amounts of calculation time, currently beyond our reach.

\section{Conclusions}\label{sconclusions}

We have presented a detailed study of the hadronic spectrum of $G_2$-QCD. 
We found that for sufficiently small quark masses a splitting of the spectrum is observed into a Goldstone sector, an intermediate bosonic sector, and a nucleonic sector, quite similar to the situation in ordinary QCD. 
The spectrum also shows strong evidence of spontaneous chiral symmetry breaking, like the emergence of the aforementioned Goldstone bosons, or the non-degeneracy of parity partners. 
Therefore, the hadronic physics appears to be qualitatively similar to QCD, even tough there are many more states in the spectrum. 
Unfortunately we could not reliably determine the mass of the lightest hybrid, though this would be crucial in assuring that the nucleon dynamics is truly similar to QCD. 
This will require a much more sophisticated spectroscopy analysis in the future.

We have also shown that the scale hierarchy of the vacuum reflects itself in the phase structure at finite densities. 
We found a number of transitions, particular for light quark mass, which correlate with the scales of the hadron spectrum. 
In fact, we found even an additional transition. This already indicates a very rich phase structure of the theory at finite densities. 
We also find some hints that a phase dominated by fermionic hadrons may exist at quark chemical potentials of about 300-600 MeV.

Besides understanding in more detail the already observed phase structure, the next logical step is to go to smaller lattice spacings. 
This would ensure that we can disentangle the transition occurring at the nucleon scale from possible lattice artifacts. 
Also, larger volumes will be necessary to reduce artifacts from the residual temperature. 
Both steps are necessary to show whether a genuine nuclear matter phase is present, which would be of central importance for a qualitative understanding of fermionic effects in finite density QCD, and eventually neutron stars.

\begin{acknowledgments}
\noindent 
This work was supported by the Helmholtz International Center for FAIR within the LOEWE initiative of the State of Hesse.
We are grateful to Jonivar Skullerud for helpful discussions. A.\ M.\ was supported by the DFG under grant number MA3935/5-1. B.\ W.\ was supported by the DFG
under grant number Wi777/11-1 and the
graduate school GRK 1523/1. L.\ v.\ S. was supported by the European Commission, FP-7-PEOPLE-2009-RG, No. 249203. Simulations were performed on the LOEWE-CSC at the University of Frankfurt and on the HPC cluster at the University of Jena.
\end{acknowledgments}

\appendix 

\renewcommand{\eprint}[1]{ \href{http://arxiv.org/abs/#1}{[arXiv:#1]}}


\bibliography{paper}


\end{document}